\documentclass[twocolumn,preprintnumbers,amsmath,superscriptaddress,amssymb]{revtex4}
\usepackage{amsmath,amssymb,comment,color}
\usepackage{graphicx}
\usepackage{dcolumn}
\usepackage{bm}

\begin{document}

\title{Topological mechanics of origami and kirigami}
\author{Bryan Gin--ge Chen}
\altaffiliation[Current address: ]{Department of Physics, University of Massachusetts, Amherst, MA  01002, USA}
\affiliation{Instituut-Lorentz, Universiteit Leiden, 2300 RA Leiden, The Netherlands}
\author{Bin Liu}
\altaffiliation[Current address: ]{ School of Natural Sciences,
University of California, Merced, CA 95343, USA}
\affiliation{Department of Physics, Cornell University, NY 14853, USA}
\author{Arthur A. Evans}
\altaffiliation[Current address: ]{Department of Mathematics, University
of Wisconsin, Madison, WI 53706, USA}
\affiliation{Department of Physics, University of Massachusetts, Amherst, MA  01002, USA}
\author{Jayson Paulose}
\affiliation{Instituut-Lorentz, Universiteit Leiden, 2300 RA Leiden, The Netherlands}
\author{Itai Cohen}
\affiliation{Department of Physics, Cornell University, NY 14853, USA}
\author{Vincenzo Vitelli}
\affiliation{Instituut-Lorentz, Universiteit Leiden, 2300 RA Leiden, The Netherlands}
\author{C.D. Santangelo}
\affiliation{Department of Physics, University of Massachusetts, Amherst, MA  01002, USA}
\date{\today}

\begin{abstract}
Origami and kirigami have emerged as potential tools for the design of mechanical
metamaterials whose properties such as curvature, Poisson's ratio, and existence of
metastable states can be tuned using purely
geometric criteria. A major obstacle to exploiting this property is
the scarcity of tools to identify and program the flexibility of fold patterns.
We exploit a recent connection between spring networks and quantum
topological states to design origami with localized folding motions at
boundaries and study them both experimentally and theoretically.
These folding motions exist due to an underlying
topological invariant rather than a local imbalance between
constraints and degrees of freedom. 
We give a simple example of a quasi-1D folding pattern that realizes such topological states.
We also demonstrate how to generalize these topological design
principles to 2D.
A striking consequence is that a
domain wall between two topologically distinct, mechanically rigid
structures is deformable even when constraints locally match the
degrees of freedom.
\end{abstract}

\maketitle

Recent interest in origami mechanisms has been spurred by advances in
fabrication and manufacturing \cite{ryu2012,tolley2014,liu2014}, as
well as a realization that folded structures can form the basis
of mechanical metamaterials
\cite{Wei2013,Schenk2013,Silverberg2014,Lv2014,metasheets}. 
The ability to identify kinematic mechanisms
-- allowable folding motions of a crease pattern --
is critical to the use of
origami to design new deployable structures and mechanical metamaterials. For example, the mechanism in the celebrated
Miura ori that allows it to furl and unfurl in a single motion
\cite{miura,Evans2015} is also the primary determinant of the fold pattern's
negative Poisson ratio \cite{Wei2013,Schenk2013}. Identifying these
mechanisms becomes more challenging when the number of apparent
constraints matches the number of degrees of freedom (DOF). 

When there is an exact balance between DOF and constraints in a periodic structure, 
the structure is marginally rigid \cite{OHern2003,Alexander1998}.
In such a case, new mechanical properties such as nonlinear response
to small perturbations emerge \cite{Wyart2005,Wyart2008,Gomez2012,Ulrich2013}.  
A recent realization is that the flexibility of such solids
may be influenced by nontrivial topology in
the phonon band structure \cite{kane2013,lubenskyreview}. Here, we show how to extend 
these topological ideas to origami and kirigami. 
We show that periodically-folded sheets may exhibit
distinct mechanical ``phases'' characterized by a topological
invariant called the {\em topological polarization}, recently
introduced by Kane and Lubensky \cite{kane2013} using a mapping
of mechanically marginal structures
to topological insulators \cite{hasan2010colloquium}.
The importance of this invariant has emerged in the study of the soft
modes of spring networks \cite{lubenskyreview}, and the nonlinear
mechanics of linkages \cite{chen2014} and buckling \cite{jaysonanne}. 
As in these examples, the
phases in our origami and kirigami structures exhibit localized
vibrational modes on certain boundaries,
and transitions from between topological phases are characterized
by the appearance of bulk modes that cost zero energy. These are the hallmarks
of topologically protected behavior in classical mechanical systems~\cite{Prodan2009,
Po2014,Xiao2015,Yang2015,Nash2015,Wang2015,Wang2015a,Susstrunk2015}. Topology
provides a new knob to tune how materials and, as we show here,
origami and kirigami structures, respond to external perturbations.

We denote by {\em origami}, mechanical structures consisting of rigid 
flat polygonal plates joined by hinges. 
We will first discuss origami with no missing plates or 
``holes'', and then generalize to {\em kirigami},
defined to be origami where such holes are allowed. 
We will consider the mechanics of origami in the geometrical limit --
folds will cost zero energy and faces do not stretch or bend.

To demonstrate the power of our approach, we introduce an example of
a 1D strip of origami
analogous to the Su-Schrieffer-Heeger polyacetylene model
\cite{kane2013,suschriefferheeger}. It
admits localized modes and stresses determined and protected by topology, which we 
realize and characterize in experiments. Additionally we show how to
generalize this to 2D
periodic origami sheets, where we have observed a striking property
that causes origami without holes to have zero
topological polarization. We give examples of hinged structures with
holes (kirigami) that
do admit distinct polarizations and thus can be used as building
blocks for metasheets with programmable local flexibility.

\begin{figure}
\begin{center}
\includegraphics[width=0.4\textwidth]{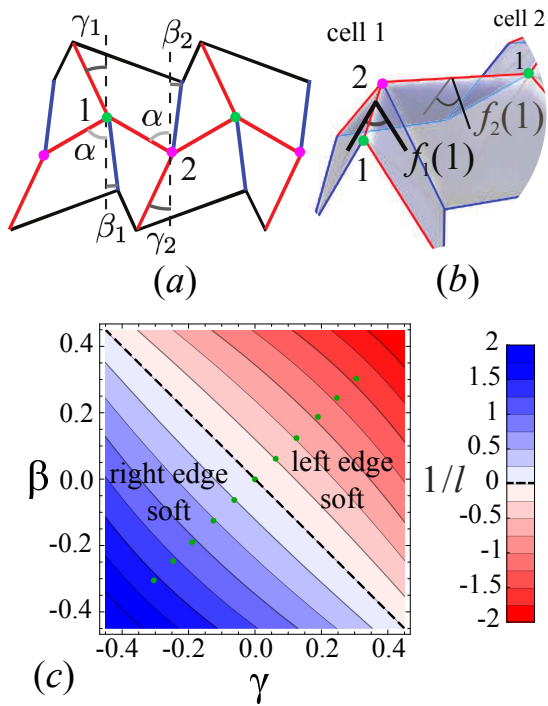}
\caption{A quasi-1D origami strip. (a) 
A unit cell of the fold pattern corresponding to the origami mechanism with
planar angles labeled.  
Red (blue) creases are mountain (valley) folds, respectively. 
(b) A 3D
depiction of a part of the strip with folding angles $f_i(n)$
labeled. (c) The phase diagram where
$\alpha=\pi/3$, $\beta\equiv\beta_1
= \beta_2$ and $\gamma\equiv\gamma_1=\gamma_2$. The colors indicate
the phase (blue for right-localized, red for left-localized); the contours
and intensity of color follow the inverse decay length $1/l$ (see legend). Configurations where
folds at a vertex become collinear lie on $\gamma=\beta$, and 
the green points along that line were constructed in experiment (along this
line, $(\gamma,\beta)$ and $(-\gamma,-\beta)$ are related by a
rotation in 3D). 
}
\label{fig1}
\end{center}
\end{figure}

{\em Quasi-1D origami strip.}--- We start with a simple quasi-1D origami structure.  Consider an origami strip of zig-zagging rigid quadrilateral plates, depicted in
Fig.~\ref{fig1}, consisting of a periodically-repeating unit cell of
two four-fold vertices. Each vertex in a cell (labeled by $n=1,2$) has
four creases (Fig.~\ref{fig1}(a)), and one 
DOF \cite{hull2012} that we track with the dihedral 
angles of the bolded crease, $f_n(j)$, where $j$ indexes the unit cell
(Fig.~\ref{fig1}(b)). 
Each adjacent pair of
dihedral angles is coupled by the kinematics of the intervening vertex. As each
vertex contributes a DOF and a constraint, this origami structure has marginal rigidity.

We analyze the mechanical response of the origami strip by determining
its configurations analytically as functions of the fold pattern angles
 $\beta_1$,$\beta_2$, $\gamma_1$ and $\gamma_2$ (defined in
Fig.~\ref{fig1}(a)). We
define a generalized displacement $u(j) = \cos f_2(j)+1$. The function
$u(j)$ encodes the dihedral angle $f_2$ of the right-most fold
of unit cell $j$, and satisfies
\begin{eqnarray}\label{eq:u}
u(j+1)= \kappa(\alpha,\beta_1,\beta_2,\gamma_1,\gamma_2) u(j),
\end{eqnarray}
where
\begin{equation}\label{eq:kappa}
\kappa= \left[\frac{\sin(\alpha-\beta_1) \sin(\alpha-\gamma_1)}{\sin(\alpha+\beta_1) \sin(\alpha+\gamma_1)}\right]\left[ \frac{\sin(\alpha-\beta_2) \sin(\alpha-\gamma_2)}{\sin(\alpha+\beta_2) \sin(\alpha+\gamma_2)}\right] .
\end{equation}
The derivation is an application of the spherical law of cosines and
is given in the SI. The fact that $u(j)$ determines $u(j+1)$ implies that the
strip has one degree of freedom, globally. 
Eq.~(\ref{eq:u}) is solved by an exponential 
$u(j)=u(0)\exp[j\ln(\kappa)]$ with deformation localized to one side
or the other, following
the sign of the inverse decay length $l^{-1}=\ln\kappa$. 

The mechanical ``phase diagram'' in Fig.~\ref{fig1}(c) shows the
values of $l^{-1}$ for
patterns with
$\gamma_1=\gamma_2\equiv\gamma,\beta_1=\beta_2\equiv\beta$. 
There are two ``phases'' distinguished by the sign of $\ln\kappa$,
which is determined here by the sign of $\gamma+\beta$, a
quantity not obviously related to any symmetry-breaking.
When $\gamma+\beta>(<)0$, $\kappa<(>)1$, and by Eq.\ (\ref{eq:u}), the mechanical 
response is localized to the left (right) of the origami strip.
A special role is played by fold patterns with $\kappa= 1$, where the
decay length diverges and $u(j)$
neither grows nor shrinks (denoted by the dashed line). This is precisely the condition for which a
global kinematic mechanism exists and the fold pattern is {\em
deployable} \cite{pellegrinobook}. As an example, when
$\gamma=\beta=0$, the strip realizes a row of the Miura ori fold
pattern, which has a single collapse motion.
More generically, however, as long as $\ln \kappa$ never
changes sign, the deformation in a strip, $u(j)$, is localized even if the values of
$\alpha, \beta_j, \gamma_j$ vary due to disorder or imperfections,
i.e.\ as long as the material remains within the same phase.
The existence of phases of robust, boundary-localized zero-energy
deformations separated by critical configurations with bulk zero modes
suggests that the origami strip has {\em topologically protected properties}.  

To make the topology explicit, we calculate a topological invariant of
the above phases. Unlike in periodic spring networks with marginal rigidity
\cite{kane2013,asimowroth,Calladine1978}, a linear analysis is
inadequate to capture the topology of the origami strip.  
Coplanar hinges in the flat state are redundant constraints, and this results in 
extra zero modes at linear order which do not extend to higher order.  In the SI, 
we derive a rigidity matrix capturing the {\em second-order} deformations of this 
structure and show that it has the same pattern of entries as the Hamiltonian of the Su-Schrieffer-Heeger chain
of Refs.\ \onlinecite{kane2013,suschriefferheeger}.
Therefore, phases of the origami strip are characterized by their topological 
polarization $\vec{P}_T$ \cite{kane2013,lubenskyreview}, defined as a winding number of
the determinant of the rigidity matrix
\footnote{For structures with $\kappa=1$, that lie on the transition
between the two distinct topological phases, the winding number and
hence polarization is not defined.}.  Indeed, the sign of $\ln\kappa$ is precisely
correlated with the topological polarization, and thus the fact that
different edges are soft or stiff in different phases is a
manifestation of the bulk-boundary correspondence
\cite{hasan2010colloquium} in this system.  
While topological modes in
1D linkages have been found to lead to propagating domain walls
\cite{chen2014,vitellipreprint}, this is not possible for our 1D
strip.  In Eq.\ (\ref{eq:kappa}), $\kappa$ depends only on the fold
pattern angles $\alpha,\beta_j,\gamma_j$, not the dihedral
angles $f_j$ -- this means that
the topological polarization of the unit cell cannot change via
the zero-energy deformations, which would be necessary for propagation.  

\begin{figure}
\begin{center}
\includegraphics[width=0.45\textwidth]{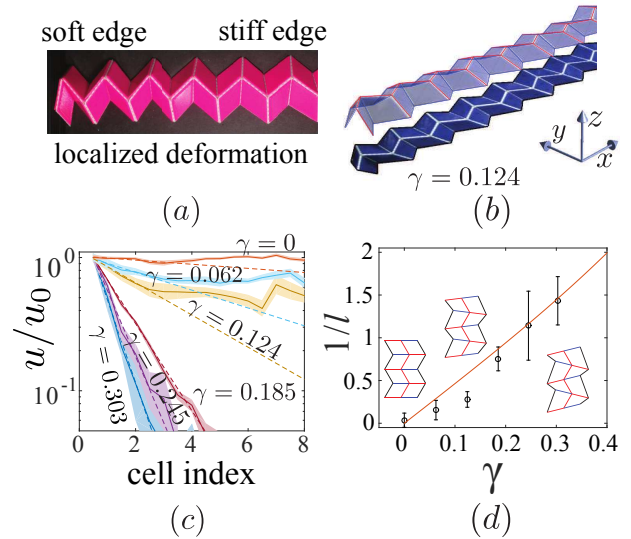}
\caption{(a) Localized 
deformations in an experimental realization of the origami strip
($\alpha=\pi/3,\gamma_1=\gamma_2=\beta_1=\beta_2=0.062$).
(b) 3D reconstruction of the
configuration of the strip from a flat image ($\gamma=0.124$). 
(c) Normalized generalized strain $u/u_0$ as a function 
of distance from the deformable boundary (measured in number of unit
cells) from experiments (shaded curves) with fits to an exponential decay
(dashed lines). Each curve shows the average of data from 
$6-30$ experimental images and has a width equal to the
standard error.
Folding angles $f_2(1)$ (related to $u_0$ via
$u_0=1+\cos f_2(1)$) at cell 1 varied from 1.06 to 1.45. 
(d) Inverse decay lengths ($1/l$) versus $\gamma$, where data points are averages
over the fitting coefficients of all images for each $\gamma$ and error bars show 
95\% confidence bounds.  The analytical result
for $l^{-1}=\ln\kappa$ (Eq.\ \ref{eq:kappa}) is plotted as a solid
red line. The deviations for
small $\gamma$ arise from a ``uniform bending mode'' (see text).
}
\label{fig2}
\end{center}
\end{figure}

To test the consequences of Eq.~(\ref{eq:u}) away from the ideal
limit, in a structure where faces can bend and hinges can twist, a mylar
sheet (200 $\mu$m thick) is perforated by a laser cutter into the desired
crease pattern, rendering it foldable along lines of perforations. We
strengthen the facets by sandwiching the mylar sheet between pairs of
$1~\textrm{mm}$ thick, plastic plates made of polylactic acid (PLA)
on a 3D printer. To mount the plastic plates onto the mylar sheet, we
use a pushed-in clip design: one facet has clips and the corresponding
facet has holes. Equivalent holes are cut on the mylar sheet so that
the clips can be pushed through to meet the holes on the plate on the
other side of the mylar. An example of the assembled origami structure
is shown in Fig.~\ref{fig2}(a).  Here, we fixed the angle $\alpha=\pi/3$ 
and varied $\gamma\equiv\beta_1=\gamma_1=\beta_2=\gamma_2$ to explore the
localization of the deformation within one phase (with $\kappa<1$)
\footnote{Note that by rotating the structure by 180$^\circ$ one
ends up with a ``new'' unit cell with $\gamma\rightarrow-\gamma$ and
$\beta\rightarrow-\beta$, which is in the opposite topological
phase.}.
A video camera captured the deformation of the strip from above
as it was symmetrically compressed. The position of each vertex
was obtained via image analysis, and fit with a 3D model to reconstruct the
complete morphology of the origami strip, as
shown in Fig.~\ref{fig2}(b). Finally, the folding angles along the interior
creases were extracted from the 3D shape and were used to compute the
generalized strain $u$.
Fig.~\ref{fig2}(c) shows the strain as a function of distance along the
strip for samples with different values of the pattern parameter
$\gamma$. Observe that there is a ``soft'' edge (cell index 0), where the deformation
is high, and on the other end a ``stiff'' edge, with low deformation.

As shown by a semi-log fit (dashed lines in Fig.~\ref{fig2}(c)), the strains
decay exponentially at small distances from the soft edge. For small $\gamma$, the folding
angles level off to a roughly constant value at larger distances, which violates
Eq.~(\ref{eq:u}).  The constant folding angle background corresponds
to the activation of a mode with uniform deformation. This mode is easy to 
excite as it is the zero energy
mode at $\gamma=0$ and thus remains very low energy for small $\gamma$. A deviation from the ideal 
geometrical limit is possible due
to the finite flexibility of the facets and the finite crease thicknesses. 
Despite the non-ideality of the experimental origami strip, the
decay lengths extracted from the fit are in good agreement with
$1/l=\ln\kappa$ (Fig.~\ref{fig2}(d)), confirming the robustness of
our topological design principle.

{\em Two-dimensional origami.}--- Having established that marginally rigid 1D periodic origami can exhibit
topological phases, we now ask whether marginality also leads to similar phases 
in 2D origami. 
We first characterize the class of {\em marginally rigid} 2D periodic
origami and show that they must have a triangulated crease pattern.
To avoid trigonometric complexity inherent to a folding angle
representation, we model the kinematics of
triangulated origami as a central-force spring
network with vertices as joints and hinges as 
springs. Triangles in such a network automatically enforce the
no-bending constraint on the facets.  Arbitrary origami can be modeled with spring
networks, but nontriangular faces require additional internal springs to
remain rigid.

In this framework, each joint has 3 degrees of freedom and each spring
adds one constraint, so marginal structures satisfy $E=3V$ where $E$ is
the number of bonds and $V$ is the number of joints. In a triangulated surface without 
a boundary, each of the $F$ faces is a triangle, so 
$3F=2E$.  The Euler characteristic $\chi$ is defined as
$\chi=V-E+F$; thus we obtain $E=3(V-\chi)$.

Periodic origami structures in 2D have the topology
of the torus and thus $\chi=0$, which shows that {\em triangulations are marginally rigid}.
While achieving marginality in granular packings and glassy networks
requires some fine-tuning in pressure or coordination, the analogous origami triangulations
arise naturally. Any non-triangular plate in an origami pattern can be
triangulated by adding diagonals, and the bending of non-triangular plates 
in real origami can be modeled as the addition of new
creases \cite{Wei2013,Silverberg2014}.  

\begin{figure}
\begin{center}
\includegraphics[width=0.5\textwidth]{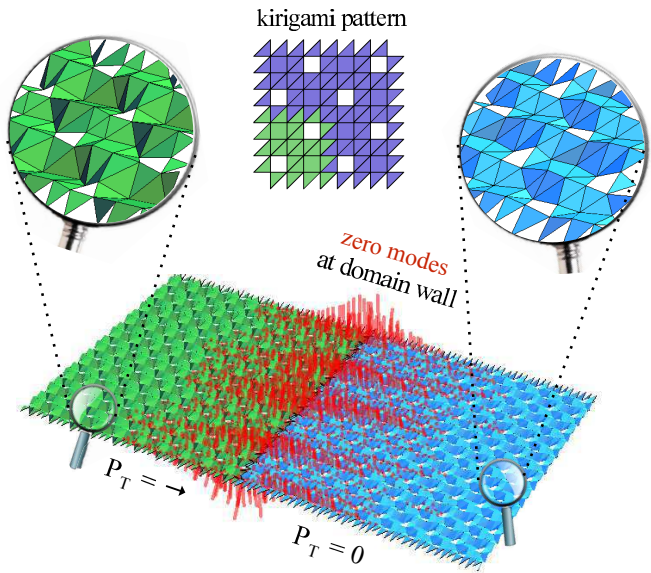}
\caption{Topologically protected zero mode (red) in a kirigami
heterostructure (left green with topological polarization
$\vec{P}_T=(1,0)$ / right blue with $\vec{P}_T=(0,0)$). 
Numerically the mode
depicted has energy nearly indistinguishable from the translation modes. This
is a close-up of a larger periodic $50\times5$ system, divided into
two $25\times5$ domains (two copies in the shorter direction are shown).  The 
magnifying
glass insets show the fine structure of four unit cells of each type, and between 
them is a schematic showing how the 
quadrilateral plates, strips of triangles, and quadrilateral holes are
joined by hinges. 
The schematic shows four unit cells, with the lower left 
cell highlighted in green. 
}
\label{fig4}
\end{center}
\end{figure}

One might now expect a variety of topological phases upon changing
the angles and lengths of a triangulated crease pattern, by
analogy with the 1D strip.  Surprisingly, our calculations indicate
otherwise.
As discussed above, an analysis of the rigidity of {\em flat} origami must
go beyond linear order. 
To bypass this complication, we consider periodic triangulated origami where 
we break the flat-state degeneracy by introducing small vertical displacements
to the vertices. 
The linear rigidity and topological properties of such a structure
can then be expressed in terms of
the (Fourier-transformed) rigidity matrix $\mathbf{R}$ for its
associated spring network \cite{kane2013,lubenskyreview}. 
However, for all triangulated periodic fold patterns we have considered, the function
$\det \mathbf{R}(\mathbf{q})$, \textit{a priori} a complex-valued
function, is in fact real-valued for all $\mathbf{q}$ in the
Brillouin zone \footnote{More precisely, since the function
$\det\mathbf{R}(\mathbf{q})$ is gauge dependent, the claim is that
this is true in the gauge arising from a ``balanced'' unit cell as
defined in Ref.\ \cite{kane2013,lubenskyreview}.}!
Though a proof of this statement for all triangulated origami eludes us, 
extensive numerical tests on a large number of distinct fold patterns bear out this
conjecture. We give details and partial results in the SI. 

A consequence of the ``reality'' property is that the winding numbers
of $\text{Arg }\det\mathbf{R}(\mathbf{q})$ along any closed curves in the
Brillouin zone must be zero (when defined), and hence the
topological polarization $\vec{P}_T$ must vanish. 
Localized boundary modes for such origami still exist, but must be isotropically 
distributed.  Even if the
hinges in a unit cell break left-right symmetry, the number of
boundary modes per unit cell on each edge of a finite patch is left-right and 
up-down symmetric.  If in fact all triangulated periodic origami structures 
have this property, the only way to get an imbalance in the number
of zero modes at the boundary of origami is by
locally adding or removing constraints.  This behavior contrasts with that of
the 1D strip of origami studied above as well as 
3D periodic networks and marginal spring networks confined to 2D.

{\em Topological kirigami.}--- Thus the question remains:
do there even exist 2D periodic hinged structures with a nonzero topological polarization?
The answer is yes, but we must go beyond origami to 
\emph{kirigami}, folded structures with holes.
There is a simple way to generate marginal kirigami
from triangulated origami.  Cutting out an adjacent pair of 
triangles removes one bond from the associated
spring network, eliminating a constraint. Likewise, merging
two triangles into a rigid quadrilateral plate adds a constraint.
We therefore modify
a triangular lattice by cutting and merging twice, resulting in a
structure with two quadrilateral plates and two quadrilateral holes
per unit cell (top center of Fig.\ \ref{fig4}).  Now
$\det\mathbf{R}(\mathbf{q})$ is complex-valued, and by randomly
perturbing a flat realization, we find the ``green'' (left) 
and ``blue'' (right) structures depicted in Fig.\ \ref{fig4}, which have
$\vec{P}_T=(1,0)$ and $(0,0)$, respectively (see SI for more details).
With free boundary conditions, the boundary soft modes in the green
kirigami are polarized to the $+x$ edge (analogous to the 1D strip and
in contrast to the blue kirigami and all triangulated origami
structures we tested).

Finding the green kirigami answers the question above positively, and
we leave a determination of the possible phases that can
occur in the modified triangular lattice to future work. A full
characterization will likely be difficult due to the
high dimensionality of the realization space (c.f.\ Ref.\
\onlinecite{weylpoints} which shows the complexity of the phase diagram in a 
simpler mechanical system). We thus switch gears and present an example of 
localized modes designed into a kirigami ``heterostructure'' to
illustrate the power of our techniques. In Fig.\ \ref{fig4}, we 
show a kirigami structure that exhibits zero modes localized at a
domain wall (one per unit cell) between the two kirigami structures 
described above. 
The zero modes render the heterostructure flexible in the vicinity of the
domain wall (the mode depicted leads to out-of-plane wrinkling), while
keeping it rigid further away. By contrast, a domain wall
between distinct patterns with equal polarization has no such localized
modes (see SI). 
In general, domain walls between structures with different polarizations
create either ``soft'' lines along which the system easily
deforms, or ``stressed'' lines which first bear the loads
under applied strains \cite{jaysonanne}. Similar effects may arise at point 
defects in otherwise uniform polarized structures \cite{paulose2015}. 

{\em Outlook.}--- We have demonstrated that origami and kirigami structures are
characterized by a topological polarization that classifies the ways
that a marginally rigid fold pattern can be floppy close to
its boundaries.  Our results give strong constraints on the types of boundary modes that
can be created in origami and will guide the design of fold patterns
that achieve a targeted mechanical response.
In the design space of geometric realizations, two
structures with different polarizations
must be separated by globally flexible, i.e.\ deployable,
realizations. Thus not only can structures with distinct phases be
combined in real space to form domain walls with useful functionality, but also
they can be used to find deployable patterns in
design space. These realization spaces are
high dimensional in general, so the problem of
determining simple rules to create a given polarization
remains open.  

We acknowledge conversations with T.C.\ Lubensky and R.D.\ Kamien. We
are grateful for funding from the NSF through EFRI ODISSEI-1240441
(AAE,BL,IC,CS), as well as financial support from FOM (BGC,JP), 
from the D-ITP consortium (JP) and a VIDI grant (VV) funded by NWO.
 
\bibliography{TopologicalOrigami}

\newpage

\setcounter{equation}{0}
\renewcommand\theequation{S\arabic{equation}}

\setcounter{figure}{0}
\renewcommand\thefigure{SI-\arabic{figure}}

\begin{center}
\textbf{Supplementary Information}
\end{center}

\noindent \textbf{Appendix A: Vertex kinematics and spherical polygons}

\begin{figure}
\begin{center}
\includegraphics[width=0.35\textwidth]{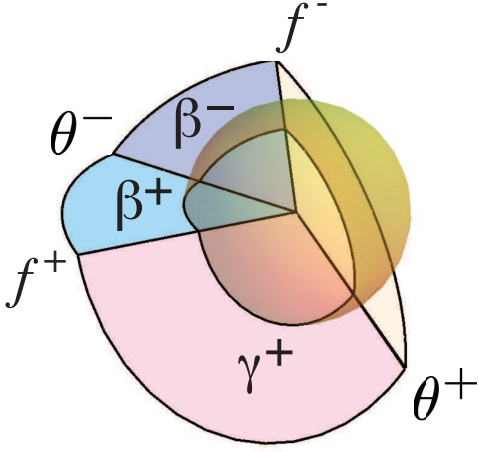}
\caption{Spherical polygons for a four-fold vertex. Angle variables
are different from those used in the main text.}
\label{fig1SI}
\end{center}
\end{figure}

The intersection between an origami vertex and a sphere traces out a
spherical polygon on the sphere surface \cite{hull2012} (Fig.\
\ref{fig1SI}). For a
four-fold vertex, this polygon is a quadrilateral whose side lengths
are inherited from the angles between adjacent folds in the fold
pattern (the unlabeled fourth angle in Fig.\ \ref{fig1SI} is $\gamma^- = 2 \pi - \beta^+ -\beta^- -\gamma^+$). The relationship between dihedral angles $f^\pm$ is
\begin{eqnarray}\label{eqSI1}
& &\cos \beta^+ \cos \gamma^+ + \sin \beta^+ \sin \gamma^+ \cos f^+ =\\
& &\cos \beta^- \cos \gamma^- + \sin \beta^- \sin \gamma^- \cos f^- \nonumber.
\end{eqnarray}
Since $\beta^+ + \gamma^+ = 2 \pi - \beta^- - \gamma^-$, Eq. (\ref{eqSI1}) becomes
\begin{eqnarray}
\sin \beta^+ \sin \gamma^+ \left(\cos f^+ + 1\right) = \sin \beta^- \sin \gamma^- \left(\cos f^- + 1\right).
\end{eqnarray}

A similar relationship between $\theta^+$ and $\theta^-$ can be derived as
\begin{eqnarray}
\sin \gamma^+ \sin \gamma^- (\cos \theta^+ + 1) = \sin \beta^+ \sin \beta^- (\cos \theta^- + 1).
\end{eqnarray}

\begin{figure}
\begin{center}
\includegraphics[width=0.35\textwidth]{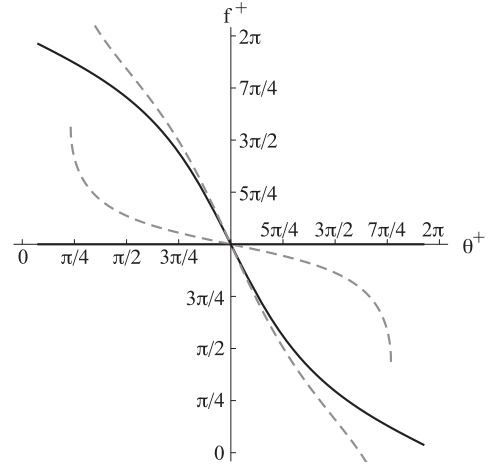}
\caption{The angle $\theta^+$ versus $f^+$ for two cases: (solid) $\beta^+ = \beta^- = \pi/3$, $\gamma^+ = \gamma^- = 2\pi/3$ and (dashed) $\beta^+ = \pi/3-\pi/10$, $\beta^- = \pi/3+\pi/10$, $\gamma^+ = \gamma^- = 2 \pi/3$. In each case, there are two distinct branches joined only at the flat state. Note that one of the branches for the solid curves lies along $f^+ = 0$.}
\label{fig2SI}
\end{center}
\end{figure}

The relationship between $f^+$ and $\theta^+$, however, is more
complex (Fig.\ \ref{fig2SI}). Define $\epsilon$ to be the angle between $\theta^+$ and $\theta^-$ given by
\begin{equation}
\cos \epsilon = \cos \beta^+ \cos \theta^+ + \sin \beta^+ \sin \theta^+ \cos f^+.
\end{equation}
and such that $\epsilon = \beta^+ + \gamma^+$ when the vertex is flat. Then
\begin{eqnarray}\label{eqSI2}
\theta^+ &=& \cos^{-1} \left(\frac{\cos \beta^+ - \cos \epsilon \cos \gamma^+}{\sin \epsilon \sin \gamma^+}\right)\\
& & \pm \cos^{-1} \left(\frac{\cos \beta^- - \cos \epsilon \cos \gamma^-}{\sin \epsilon \sin \gamma^-}\right).\nonumber
\end{eqnarray}
Either sign in Eq.\ (\ref{eqSI2}) satisfies the geometric constraint
equations; this ambiguity arises from the different branches of
$\cos^{-1}$. Yet, the branches are physical, and correspond to
different ways of choosing which folds are ``mountain'' and which
folds are ``valley'' folds. In particular, there is a branch in which
$f^+$ and $f^-$ are both either mountain or valley folds and one in
which they have the opposite sense \cite{metasheets}. In either case,
however, $\cos f^\pm$ remains the same and the exponential solution
for $u=1+\cos f$ in Eqs.\ 1 and 2 of the main text is valid.

One special case bears additional analysis. When the vertex has $\gamma^\pm + \beta^\pm = \pi$, the folds $\theta^+$ and $\theta^-$ are collinear. In that case, one branch of the vertex has both $f^+$ and $f^-$ as mountain (valley) folds. For the other branch, however, $f^+ = f^- = \pi$ is strictly satisfied and $\theta^+ = \theta^-$. That is simply the case that a vertex is folded along the line shared by $\theta^+$ and $\theta^-$, and this prevents the two adjacent folds $f^\pm$ from folding at all. This motivates the choice of variables discussed in the main text, for which $\beta^\pm = \alpha \mp \beta$ and $\gamma^\pm = 2 \pi - \alpha \mp \gamma$. Then $\beta = \gamma$ precisely when $\theta^\pm$ are collinear.

\noindent \textbf{Appendix B: Second-order rigidity matrix for 1D strip}

We label each unit cell with an index $j$ and the two vertices within a unit cell with $n=1$, $2$. For convenience, we also express the dihedral angles in terms of their difference from $\pi$, $\delta f^\pm_n(j) = f^\pm_n(j) - \pi$ for the folds $f^\pm$ of
vertex $n$ within unit cell $j$. Each vertex has one DOF, which we will parameterize with $\delta \theta^+_n(j)$. Since there is a two vertex unit cell, we combine these parameters into two vectors,
\begin{equation}
\mathbf{f}(j) = \left(\begin{array}{c}
f^-_1(j)\\
f^+_1(j)\\
f^-_2(j)\\
f^+_2(j))\end{array}\right),
\end{equation}
and
\begin{equation}
\mathbf{s}(j) = \left(\begin{array}{c}
\theta^+_1(j)\\
\theta^+_2(j)
\end{array}\right).
\end{equation}
Using spherical trigonometry, we compute $\mathbf{f}(j) = \mathbf{J} \mathbf{s}(j)$, where
\begin{equation}
\mathbf{J} = \left(\begin{array}{cc}
A^-_1 + \sigma_1 B^-_1& 0\\
A^+_1 + \sigma_1 B^+_1& 0\\
0 & A^-_2+\sigma_2 B^-_2\\
0 & A^+_2+\sigma_2 B^+_2
\end{array}\right), 
\end{equation}
$\sigma_n = \pm 1$ is an arbitrary choice of branch in configuration space the vertex should be folded,
\begin{equation}
A^\pm_n = \frac{\sin(\alpha \mp \gamma_n)}{\sin (2 \alpha)},
\end{equation}
and
\begin{equation}
B^\pm_n = \frac{1}{\sin(2 \alpha)} \sqrt{\frac{\sin(\alpha-\gamma_n) \sin(\alpha+\gamma_n) \sin(\alpha \mp \beta_n)}{\sin(\alpha \pm \beta_n)}}.
\end{equation}
For this paper, we have always assumed that $\sigma_n=1$.

To complete our description, we must enforce the linear constraints
$\delta f_1^-(j+1)-\delta f_2^+(j) = 0$ and $\delta f_2^-(j)-\delta
f_1^+(j)=0$. In Fourier space, these constraints lead to an equation of the form
\begin{equation}
\mathbf{R}(q) \mathbf{s}(q)= \mathbf{0}.
\end{equation}
for the infinitesimal deformations of the origami strip,
where
\begin{equation}
\mathbf{R}(q) = \left(\begin{array}{cc}
(A_1^- + B_1^-) e^{i q} & -(A_2^+ + B_2^+)\\
A_2^- + B_2^- & (A_1^+ + B_1^+)
\end{array}\right).
\end{equation}

\noindent \textbf{Appendix C: Real-valuedness of the determinant: a hidden symmetry}

\begin{figure}
\begin{center}
\includegraphics[width=0.45\textwidth]{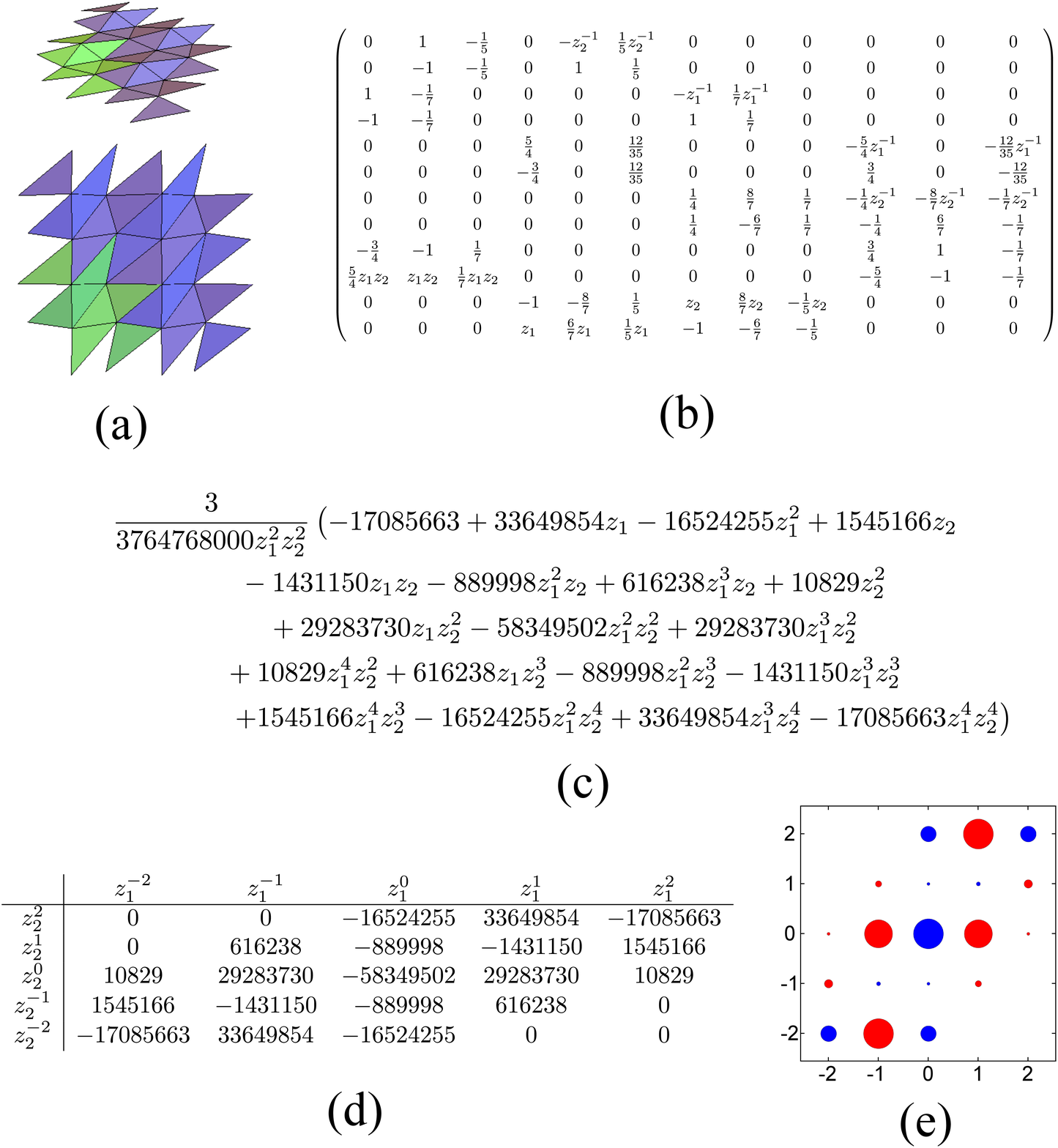}
\caption{Illustration of the reality property for triangulated
origami.  (a) A periodic triangulation based on the
triangular lattice with four vertices, 12 edges and 8 triangles. The
vertex coordinates are $(0, 0, 0)$, $(0, 1, 1/5)$, $(1, 1/7, 0)$, 
$(3/4, 1, -1/7)$ and the lattice primitive vectors are $(2,0,0)$ and
$(0,2,0)$.  Above is a side view and below is a top view.  Four unit 
cells are shown, with one highlighted in green.
(b) The Fourier transformed rigidity matrix $\mathbf{R}(z_1,z_2)$,
where $z_1=e^{iq_1a_1}$ and $z_2=e^{iq_2a_2}$. Every row corresponds
to an edge in the triangulation.
(c) The Laurent polynomial $\det\mathbf{R}(z_1,z_2)$. (d) The table of
coefficients of $\det\mathbf{R}(z_1,z_2)$, organized by powers of
$z_1$ and $z_2$.  The number
$3/3764768000$ has been factored out of all entries.  (e) A graphical
representation of the coefficients of $\det\mathbf{R}(z_1,z_2)$.  Red
dots are positive coefficients, blue dots are negative, and the
area of the disks is propoertional to the absolute value of the
coefficients (with a minimum disk size imposed for visibility). }
\label{fig:reality}
\end{center}
\end{figure}

We will restate the observed ``reality'' property here more formally.
For all periodic triangulations we have checked, there exists a choice
of unit cell so that the determinant of
the rigidity matrix $\det\mathbf{R}(\vec{q})$ is real
valued for all $\vec{q}$ in the Brillouin zone. The consequence of
this is that with such a choice of unit cell, the phase of
$\det\mathbf{R}(\vec{q})$ cannot wind and so the topological
polarization must vanish.  There are also obvious consequences for the
set of bulk zero modes which we will not discuss here.
In Fig.~\ref{fig:reality} we display a sample calculation and
visualization of this propety for a small periodic triangulated
origami, represented by a spring network.

This real-valuedness can be
expressed in the following equivalent way.  First, define $z_1=e^{iq_1a_1}$ and
$z_2=e^{iq_2a_2}$, and observe that $\det\mathbf{R}(z_1,z_2)$ is a
Laurent polynomial in those variables.  Next, note that
$\det\mathbf{R}(\vec{q})$ is real if and only if:
\begin{align*}
\det\mathbf{R}(\vec{q})&=\overline{\det\mathbf{R}(\vec{q})}\\
\det\mathbf{R}(z_1,z_2)&=\overline{\det\mathbf{R}(z_1,z_2)}\\
&=\det\mathbf{R}(\bar z_1,\bar z_2)\\
&=\det\mathbf{R}(z_1^{-1},\bar z_2^{-1}).
\end{align*}
In the third line, we use the fact that the coefficients of this
polynomial are real, and in the final line, 
we use the fact that all $\vec{q}$ in the Brillouin zone,
$z_1$ and $z_2$ lie on the unit circle and so $\bar{z_i}=z^{-1}_i$.
This shows that for all $p,q$, the coefficients of the $z_1^pz_2^q$ and
$z_1^{-p}z_2^{-q}$ terms must be equal in order for this polynomial to
be real-valued on the unit circle.

A change of unit cell merely changes $\det\mathbf{R}$ by a factor $z_1^mz_2^n$ 
where $m$ and $n$ are integers related to the change in the number of
bonds crossing unit cell boundaries \cite{kane2013}. If one imagines
the coefficients of $\det\mathbf{R}$ as a table (Fig.\
\ref{fig:reality}(d)), then
such a factor does not change the values in the table, but merely
shifts the positions of the entries in the table by a vector $(m,n)$.
Therefore, the ``reality property'' is equivalent to the 
polynomial $\det\mathbf{R}(z_1,z_2)$ being ``2D palindromic'', i.e.
\begin{equation}
\label{eq:palindromic}
\det\mathbf{R}(z_1^{-1},z_2^{-1})=z_1^kz_2^l\det\mathbf{R}(z_1^{-1},z_2^{-1}),
\end{equation}
for some integers $k,l$.
Equivalently, the table of coefficients of $\det\mathbf{R}(z_1,z_2)$
is symmetric under inversion through some point $(k/2,l/2)$.

\noindent {\em Examples and tests performed}:
We generated small triangulations of a torus via flipping edges from
the triangular lattice and in all cases checked found that the the
polynomial $\det\mathbf{R}(z_1,z_2)$ was a palindrome. To ensure that
loss of precision from floating-point operations were not an issue, we
did these computations in rational arithmetic in the Mathematica 
computer algebra system. 

\begin{figure}
\begin{center}
\includegraphics[width=0.45\textwidth]{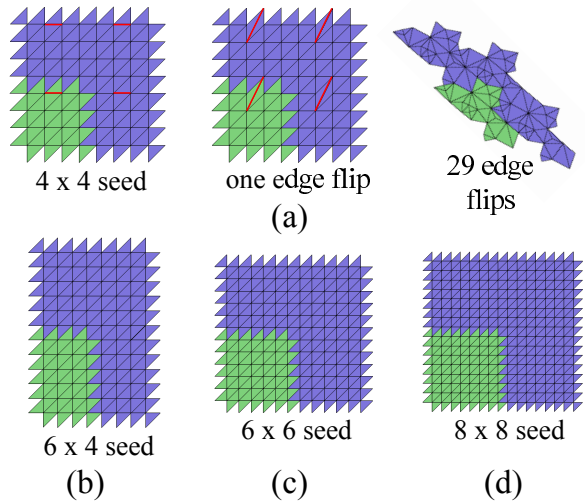}
\caption{Seed triangulations and edge flips. In each image, 
four unit cells are shown, with one highlighted in green. The 
triangulations are only shown with planar embeddings for clarity, the
actual tested triangulations were deformed into 3D with random
perturbations.
(a) left: The 4 by 4 seed triangulation with one highlighted edge, 
center: the triangulation resulting from the edge flip of the red
edge, right: a triangulation resulting from 29 random edge flips on
the seed.
(b) The 6 by 4 seed triangulation. (c) The
6 by 6 seed triangulation.  (d) The 8 by 8 seed triangulation. }
\label{fig:triangulations}
\end{center}
\end{figure}

Specifically, we began with periodic
triangulation ``seeds'' displayed in Fig.\ \ref{fig:triangulations} and performed random 
edge flips (transforming two adjacent triangles into two different
triangles on the same four vertices) repeatedly onto randomly chosen
edges. This procedure is illustrated in Fig.\
\ref{fig:triangulations}(a).

For every such crease pattern generated, we used vertex coordinates
made by randomly perturbing a flat embedding of the seed
triangulation.  These were rounded to the nearest rational number with
denominator lower than 10. The Laurent polynomial
$\det\mathbf{R}(z_1,z_2)$ was then calculated exactly and the
coefficients of different terms were compared to check whether the
polynomial was indeed palindromic as in Eq.\ (\ref{eq:palindromic}).
In some cases, a flip would cause this polynomial to vanish (due e.g.\
to a degenerate edge connecting a vertex to itself). Afterwards, it is
quite likely that further flips would produce triangulations with zero
determinant.  While the zero
polynomial is of course real-valued, if this occurred, 
in order to get more interesting tests, we rejected this triangulation and 
flipped another edge instead.  Regardless, for all triangulated origami 
checked we found that Eq.\ (\ref{eq:palindromic}) was always satisfied.  
In Table \ref{SItable} we show a summary of the tests 
performed.

\begin{table}
\centering
\begin{tabular}{ccccc}
seed   & $|V|$ & $|E|$ & $|T|$ & (runs, flips)\\
\hline
4 by 4 & 16     & 48    & 32     & (100, 1000) \\
6 by 4 & 24     & 72    & 48    & (50, 500) \\
6 by 6 & 36     & 108    & 72    & (20, 200) \\
8 by 8 & 64    & 192    & 128    & (10, 100)
\end{tabular}
\caption{Summary of tests performed of the ``reality'' property 
on different families of periodic triangulated origami. 
Each line of the table corresponds to a different seed
(depicted in Fig.\ \ref{fig:triangulations}), and displays the number of vertices $|V|$,
edges $|E|$ and triangles $|T|$ in the seed and shows the number of runs started as
well as the number of random flip operations performed. Note that a
flip preserves $|V|$, $|E|$ and $|T|$, and that the computations
require computing determinants of symbolic $|E|\times3|V|$ matrices. 
In all runs, every single triangulation generated had the reality 
property, as described in the text.}
\label{SItable}
\end{table}

\noindent {\em Partial result}: 
We can show that subdividing a triangle in triangulated origami
preserves the property that $\det\mathbf{R}(\vec{q})$ remains real. To
see this, note that such a subdivision creates one new vertex and
three edges.  
With an appropriate choice of unit cell we can ensure that this vertex
is not on the boundary and thus does not pick up factors of
$e^{iq_1a_1}$ or $e^{iq_2a_2}$.

The new vertex introduces three columns into
$\mathbf{R}(\vec{q})$ whose only nonzero entries can be in the rows
corresponding to the three new edges.  By reordering the vertices and
columns, one finds that the determinant takes a block upper-triangular form with a
$3\times3$ block on the diagonal from the new additions (which we call
$\mathbf{R}_n(\vec{q})$) as well as $\mathbf{R}_0(\vec{q})$ as another
block on the diagonal. 
By a well-known property of the determinant, this implies that
$\det\mathbf{R}(\vec{q})=\det\mathbf{R}_n(\vec{q})\det\mathbf{R}_0(\vec{q})$.

If we were able to prove an analogous result for edge flips, then we
would have shown that the palindromic property holds for all
triangulations.

\noindent \textbf{Appendix D: On the design of topological kirigami}

\begin{figure}
\begin{center}
\includegraphics[width=0.45\textwidth]{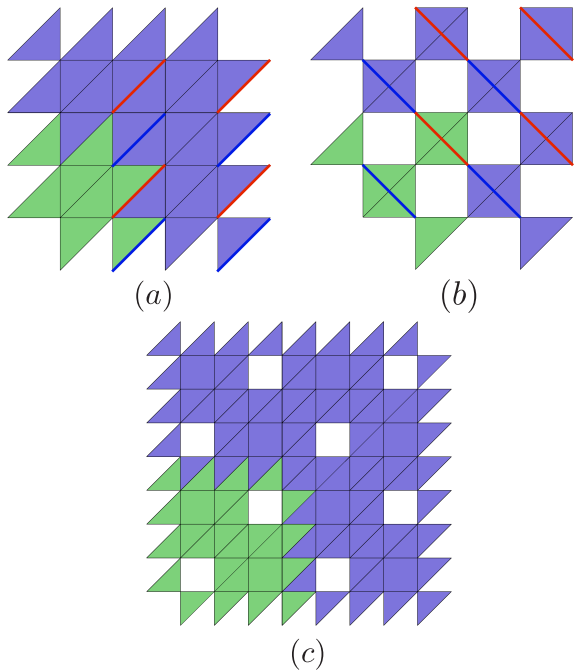}
\caption{Schematics of some structures created while designing the
kirigami. In each image, four unit cells are shown, with one
highlighted in green. These images only show the pattern of hinges and
plates, and do not represent 3D embeddings.
(a) A triangulated origami based on the triangular lattice.
Two families of creases are highlighted in red and blue which are
moved to create the next structure.
(b) The checkerboard lattice of tetrahedra.  This structure consists
of tetrahedra joined by universal joints at their vertices. (c) The
final kirigami structure.  Created by placing rows and columns of
additional triangles between tetrahedra, and then replacing the
tetrahedra with flat quadrilateral plates. }
\label{fig:design}
\end{center}
\end{figure}

\begin{figure}
\begin{center}
\includegraphics[width=0.45\textwidth]{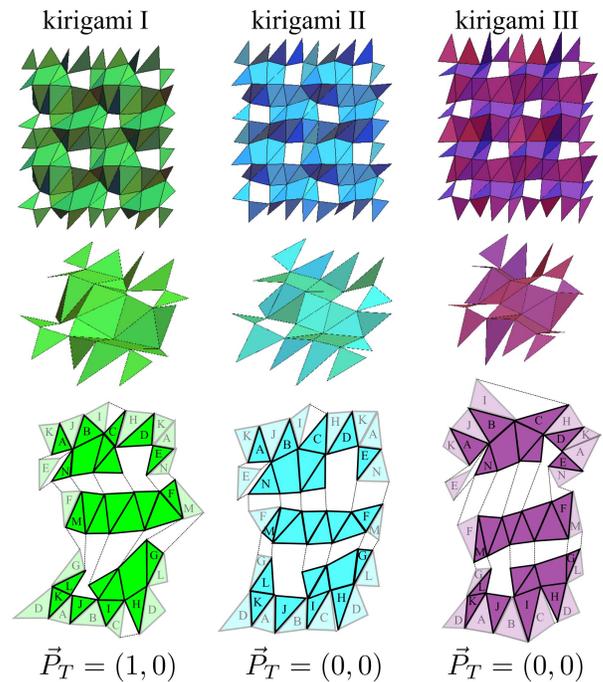}
\caption{Three kirigami unit cells with different topological
polarizations. A top view and side view of the 3D unit cells are shown 
above exploded versions that have been cut into 
strips and then flattened into the plane. The dashed lines indicate
how the strips are to be
joined to create the 3D structure shown above and the faces labeled by
letters indicate the periodicity of the structure. 
}
\label{fig3}
\end{center}
\end{figure}

We found topologically polarized kirigami via the following steps.  We
first constructed the pattern of plates and hinges as follows.
As in Appendix C, we began with a periodic spring network with 
the combinatorics of the standard triangular lattice (drawn in 
Fig.\ \ref{fig:design}(a) as a square lattice 
with all NE diagonals added). This pattern models a triangulated origami
structure, which seems to have a real determinant.  To try to break
this symmetry we attempted to add holes in the unit cell.

To add holes without changing the marginal rigidity property, we removed
one NE diagonal bond from a square $A$ and placed it in an adjacent (already
occupied) square $B$ as a NW diagonal bond.  This yields a hole in $A$
and a new tetrahedron in $B$.  If one does this for all squares in a
checkerboard pattern, one ends up with the so-called checkerboard
lattice of tetrahedra (Fig.\ \ref{fig:design}(b)).  This is the
simplest 2D periodic structure in 3D
which admits geometric realizations with distinct topological
polarizations. However, the tetrahedra are 
joined at vertices, not along hinges, so it is neither origami nor
kirigami.

To remedy this, we may add rows and columns of triangulated strips
between the holes and the tetrahedra.  Now we do have a
structure made of solid bodies joined along hinges, but the bodies are
tetrahedra and not yet flat plates. Flat plates can be modeled by a spring network 
with pyramids. By adding an additional 
vertex above the plate and attaching it to the four vertices of the we
create a minimally rigid body which is attached to the structure by
four coplanar hinges. The final pattern of hinges and plates is
shown in Fig.\ \ref{fig:design}(c) (pyramids not shown for clarity).

Given this design, we still have to find a suitable realization of
this structure as a 3D object.  We did this by beginning with a flat
embedding of a unit cell and perturbing the vertices randomly, subject 
to the condition that the quadrilteral plates remain planar.  These
randomly generated unit cells were then checked to determine if they
had a topological polarization.
Fig.~\ref{fig3} shows the unit cells of three kirigami structures with this crease
pattern but with differing topological polarizations: kirigami I has
polaarization $(1,0)$, and kirigami II and III have polarization
$(0,0)$. Their 3D structures are also displayed in an ``exploded'' view,
where they have been
cut apart into flattened strips which must be bent in spance and then glued 
together to form the kirigami structures.  Unit cells I and II were used to construct the
domain wall in Fig.~3 in the main text.

\begin{figure*}
\begin{center}
\includegraphics[width=0.9\textwidth]{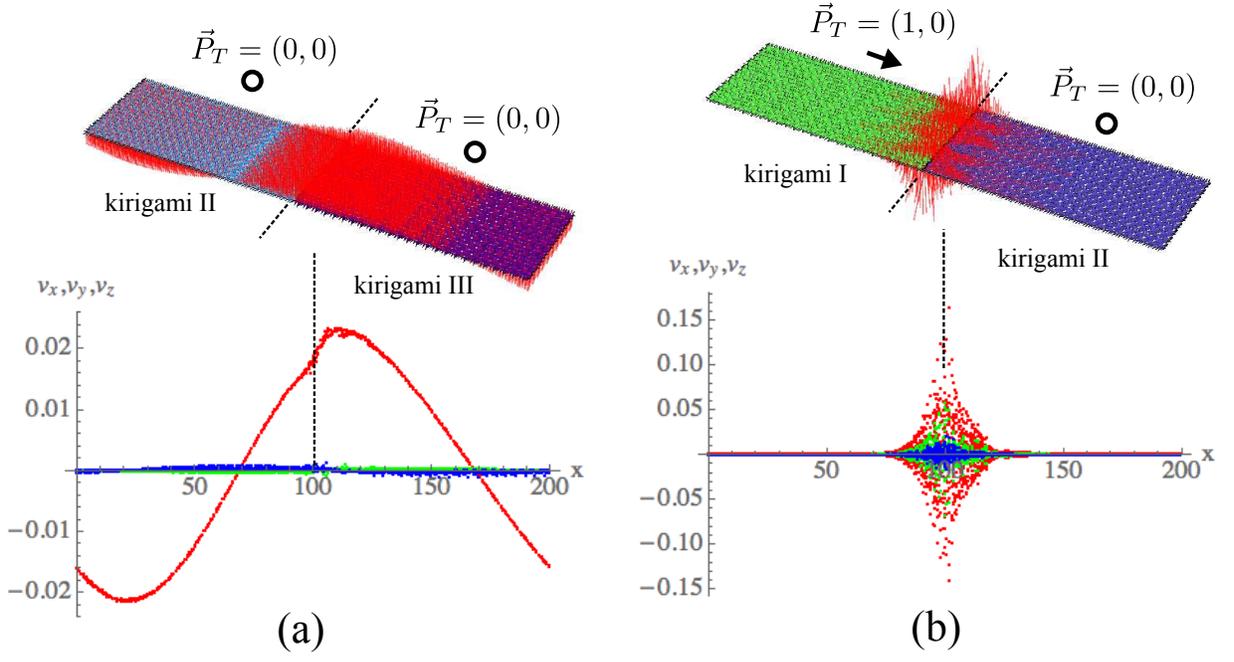}
\caption{Composite kirigami systems constructed from unit cells I, II
and III from Fig.~\ref{fig3} and their lowest energy eigenmodes. Both
structures depicted consist of 50 unit cells (along the $x$-direction)
by 5 unit cells (along the $y$-direction), which are divided into 
two domains each of size 25 by 5.
Two copies in the shorter direction are shown in the upper figures,
and the displacements in the mode are overlaid as red lines.  The 
structures are periodic in both directions, so there are domain walls
at $x=0=200$ and $x=100$.  The domain wall at $x=100$ is marked with 
dashed lines.  The lower plots show the $x$-, $y$- and
$z$-components (in blue, green, and red,
respectively) of the displacement vectors of all vertices as a
function of the $x$-positions.  (a) Composite structure of kirigami II and
III, both of which have polarization (0,0).  The lowest eigenmode has
eigenvalue $\omega^2=1.82\times10^{-9}$ and resembles a bulk bending
mode. (b) Composite structure of kirigami I (polarization (1,0)) and
II (polarization (0,0)).  The lowest eigenmode has
eigenvalue $\omega^2=1.42\times10^{-16}$ and is highly localized near
the domain wall at $x=100$.}
\label{fig:silowmode}
\end{center}
\end{figure*}

In Fig.~\ref{fig:silowmode} we show kirigami structures consisting of
pairs of unit cells from Fig.~\ref{fig3}. The domain walls do not contain any
extra or missing bonds, so there are no differences in the local
counting of degrees of freedom.
These kirigami structures are rectangular sheets with periodic boundary 
conditions in both directions, so
one might expect the lowest energy vibrational modes to be bulk bending
modes.  This is indeed the case when the two unit cells used have the
same polarization vector (Fig.~\ref{fig:silowmode}(a)).  However, when a 
structure is made of unit
cells with two distinct polarizations, one of the domain walls carries
localized very low-energy modes (Fig.~\ref{fig:silowmode}(b) and
Fig.~3 in the main text).  The eigenvalue corresponding
to the energy of the topological mode is
$\omega^2=1.42\times10^{-16}$ (comparable to the eigenvalues
corresponding to the translational zero modes), whereas the eigenvalue
corresponding to the bending mode is much larger,
$\omega^2=1.82\times10^{-9}$.

\end{document}